\documentclass[11pt]{article}

\advance\textwidth 24mm  \advance\oddsidemargin -12mm
\advance\textheight 25mm \advance\topmargin -15mm

\usepackage{amsmath, amsthm, amssymb}
\usepackage{epic,eepic}
\usepackage[dvips]{graphicx}
\usepackage[dvips]{graphics}
\usepackage[dvips]{color}
\newtheorem{theorem}{Theorem}
\newtheorem{lemma}{Lemma}
\newtheorem{proposition}{Proposition}
\theoremstyle{remark}

\def\a{\alpha}
\def\b{\beta}
\def\g{\gamma}
\def\d{\delta}

\def\sn{{\rm sn}}
\def\cn{{\rm cn}}
\def\dn{{\rm dn}}

\begin{document}

\title{On the Lagrangian structure of integrable quad-equations}

\author{Alexander I. Bobenko\thanks{Institut f\"ur Mathematik, MA 8-3,
TU Berlin, Str. des 17. Juni 136, 10623 Berlin, Germany; e-mail:
{\tt bobenko@math.tu-berlin.de}; partially supported by the DFG
Research Unit ``Polyhedral Surfaces''.} \ and Yuri B.
Suris\thanks{Institut f\"ur Mathematik, MA 7-2, TU Berlin, Str.
des 17. Juni 136, 10623 Berlin, Germany; e-mail: {\tt
suris@math.tu-berlin.de}.}}

\maketitle

\begin{abstract}
The new idea of flip invariance of action functionals in
multidimensional lattices was recently highlighted as a key
feature of discrete integrable systems. Flip invariance was proved
for several particular cases of integrable quad-equations by
Bazhanov, Mangazeev and Sergeev and by Lobb and Nijhoff. We
provide a simple and case-independent proof for all integrable
quad-equations. Moreover, we find a new relation for Lagrangians
within one elementary quadrilateral which seems to be a
fundamental building block of the various versions of flip
invariance.
\end{abstract}

\section{Introduction}
This paper deals with some aspects of the variational (Lagrangian)
structure of integrable systems on quad-graphs (planar graphs with
quadrilateral faces), which serve as discretizations of integrable
PDEs with a two-dimensional space-time \cite{BS, ABS}. We identify
integrability of such systems with their multidimensional
consistency  \cite{BS, Nijhoff}. This property was used in
\cite{ABS} to classify integrable systems on quad-graphs. That
paper also introduced a Lagrangian formulation for them. The
variational structure of discrete integrable systems is a topic
which receives increasing attention in the recent years
\cite{MPS98, Suris}, after the pioneering work
\cite{MoserVeselov}.

Lobb and Nijhoff \cite{Lobb} introduced the new idea to extend the
action functional of \cite{ABS} to a multidimensional lattice. The
key property that makes this meaningful is the invariance of the
action under elementary 3D flips of 2D quad-surfaces in $\mathbb
Z^m$. This property was established in \cite{Lobb} for several
particular cases of integrable equations. The proof involves
computations based in particular on properties of the dilogarithm
function. In the present paper, we prove the flip invariance for
all integrable quad-equations classified in \cite{ABS}; our proof
is case-independent. Note that three-dimensional discrete
integrable systems also possess Lagrangian formulations
\cite{BMS2,Lobb2}, and the flip invariance of action for the
discrete KP equation was established in \cite{Lobb2}.

A closely related version of flip invariance of action for
discrete systems of Laplace type was discussed earlier for one
concrete example by Bazhanov, Mangazeev and Sergeev \cite{BMS}.
The action functional in this paper describes circle patterns and
was introduced in \cite{BS04}. In \cite{BMS}, this action was
derived as a quasi-classical limit of the partition function of an
integrable quantum model investigated in \cite {FV} (the
Lagrangians being the quasi-classical limit of the Boltzmann
weights). Invariance of the partition function under star-triangle
transformations is a hallmark of integrability in the quantum
context, it is usually established with the help of the quantum
Yang-Baxter relation \cite{Baxter}. It is surprising that only now
a correct classical counterpart comes to the light. Here, we
extend the quasi-classical result of \cite{BMS} to the whole class
of integrable quad-equations. Finding the quantum version of our
contribution remains an open problem.

The structure of the paper is as follows. In Section \ref{sect:int
quad}, we recall the definition and the classification of
integrable systems on quad-graphs, the so-called ABS list
\cite{ABS} (see also the recent monograph \cite{the Book}). In
Section \ref{sect: 3leg}, we recall our main technical device
which plays a prominent role in the subject of the present paper,
namely the three-leg form of a quad-equation. Further, we recall a
variational (Lagrangian) interpretation of integrable
quad-equations,  again following \cite{ABS}. In Section
\ref{sect:one quad}, we prove a novel relation for Lagrangians
within one elementary quadrilateral which seems to be a
fundamental building block of the various versions of the flip
invariance. Finally, Section \ref{sect: flip} contains
generalizations of the flip invariance results from \cite{Lobb}
and \cite{BMS} with a new case-independent proof.

The flip invariance of the action functional in multidimensional
lattices is a fascinating new idea which will definitely have a
serious impact on the theory of discrete integrable systems.

\section{Integrable systems on quad-graphs}
\label{sect:int quad}

We consider systems on quad-graphs, i.e., collections of equations
on elementary quadrilaterals of the type
\begin{equation}\label{system}
    Q(x,u,y,v;\a,\b)=0,
\end{equation}
where $x,u,y,v\in\mathbb{CP}^1$ are the complex variables
(``fields'') assigned to the four vertices of the quadrilateral,
and the parameters $\a,\b\in\mathbb C$ are assigned to its edges,
as shown on Fig.~\ref{Fig:quadrilateral}.
%------------------------------------------------------------------
\begin{figure}[htbp]
    \setlength{\unitlength}{0.05em}
\begin{minipage}[t]{170pt}
\begin{center}
\setlength{\unitlength}{0.06em}
\begin{picture}(200,140)(-50,-20)
  \put(100,  0){\circle{8}} \put(0  ,100){\circle{8}}
  \put(0,0){\circle*{8}}  \put(100,100){\circle*{8}}
  \path(4,0)(96,0)\path(100,4)(100,96)
  \path(96,100)(4,100)\path(0,96)(0,4)
  \put(-10,-13){$x$}
  \put(99,-13){$u$}
  \put(99,110){$y$}
  \put(-10,110){$v$}
  \put(47,-13){$\a$}
  \put(47,105){$\a$}
  \put(-13,47){$\b$}
  \put(105,47){$\b$}
  \put(45,45){$Q$}
\end{picture}
\caption{An elementary quadrilateral}\label{Fig:quadrilateral}
\end{center}
\end{minipage}\hfill
\begin{minipage}[t]{160pt}
\begin{center}
\setlength{\unitlength}{0.06em}
\begin{picture}(200,140)(-40,-20)
  \put(100,  0){\circle{8}} \put(0  ,100){\circle{8}}
  \put(  0,  0){\circle*{8}}  \put(100,100){\circle*{8}}
  \path(4,0)(96,0)
  \path(0,4)(0,96)
  \path(2.8,2.8)(97.2,97.2)
  \put(-10,-13){$x$}
  \put(99,-13){$u$}
  \put(99,110){$y$}
  \put(-10,110){$v$}
  \put(47,-13){$\psi$}
  \put(-13,47){$\psi$}
  \put(57,47){$\phi$}
  \put(10,5){$\a$}\put(80,90){$\a$}
  \put(3,14){$\b$}\put(90,78){$\b$}
\end{picture}
\caption{Three-leg form of a quad-equation.}\label{Fig: 3leg}
\end{center}
\end{minipage}
    \end{figure}
%------------------------------------------------------------------------
It is required that opposite edges of any quadrilateral carry the
same parameter. The function $Q$ is assumed to be multi-affine,
i.e., a polynomial of degree one in each field variable. Moreover,
it is supposed to possess the following property:
\begin{itemize}
\item \textit{Symmetry}: The equation $Q=0$ is invariant under the
dihedral group $D_4$ of the square symmetries:
\begin{eqnarray*}\label{symmetry}
Q(x,u,y,v;\a,\b)=\varepsilon_1 Q(x,v,y,u;\b,\a)= \varepsilon_2
Q(u,x,v,y;\a,\b),
\end{eqnarray*}
with $\varepsilon_1,\varepsilon_2=\pm 1$.
\end{itemize}

\begin{figure}[htbp]
\begin{center}
\setlength{\unitlength}{0.045em}
\begin{picture}(200,250)(0,-10)
 \put(0,0){\circle*{10}}    \put(150,0){\circle{10}}
 \put(0,150){\circle{10}}   \put(150,150){\circle*{10}}
 \put(70,200){\circle*{10}} \put(220,200){\circle{10}}
 \put(70,50){\circle{10}}   \put(220,50){\circle*{10}}
 \path(0,0)(145,0)       \path(0,0)(0,145)
 \path(150,5)(150,150)   \path(5,150)(150,150)
 \path(3.53,153.53)(70,200)    \path(150,150)(216.47,196.47)
 \path(70,200)(215,200)
 \path(220,195)(220,50) \path(220,50)(153.53,3.53)
 \dashline{8}(0,0)(66.47,46.47)
 \dashline{8}(70,55)(70,200)
 \dashline{8}(75,50)(220,50)
 %\dashline{3}(150,0)(0,150)
 %\dashline{3}(150,0)(70,50)
 %\dashline{3}(70,50)(0,150)
 %\dashline{3}(150,0)(220,200)
 %\dashline{3}(0,150)(220,200)
 %\dashline{3}(70,50)(220,200)
 \put(-25,-15){$x$} \put(-30,145){$x_3$}
 \put(162,-15){$x_1$} \put(160,140){$x_{13}$}
 \put(235,45){$x_{12}$} \put(230,210){$x_{123}$}
 \put(70,30){$x_2$}  \put(35,210){$x_{23}$}
 \put(70,-15){$\a_1$}   \put(110,57){$\a_1$}
 \put(80,158){$\a_1$}   \put(120,208){$\a_1$}
 \put(13,32){$\a_2$}    \put(190,15){$\a_2$}
 \put(5,177){$\a_2$}    \put(190,165){$\a_2$}
 \put(-27,75){$\a_3$}\put(157,75){$\a_3$}
 \put(44,112){$\a_3$}\put(227,115){$\a_3$}
\end{picture}
\caption{Three-dimensional consistency}\label{3D}
\end{center}
\end{figure}
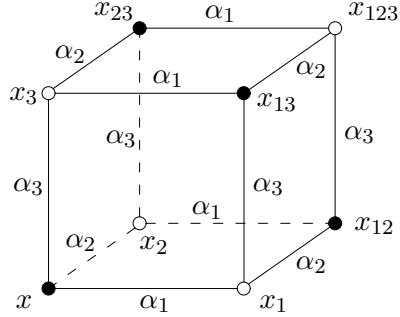

As in \cite{BS, Nijhoff}, we consider integrability as synonymous
with 3D consistency. Recall that equation (\ref{system}) is called
{\em 3D-consistent} if it may be consistently imposed on a
three-dimensional lattice, so that one and the same equation hold
for all six faces of any elementary cube (up to the parameter
values: it is supposed that all edges of each coordinate direction
carry their own parameter). More precisely, initial data $x, x_1,
x_2, x_3$ determine uniquely the values $x_{12}, x_{13}, x_{23}$
by means of the equations on the faces adjacent to the vertex $x$.
After that, one has three different equations for $x_{123}$,
coming from the three faces of the cube adjacent to this vertex,
see Fig.~\ref{3D}. Now 3D consistency means that these three
(\textit{a priori} different) values for $x_{123}$ coincide for
any choice of the initial data $x,x_1,x_2,x_3$.

%-----------------------------------------------------------------
%-----------------------------------------------------------------
Integrable equations on quad-graphs with multi-affine and
$D_4$-symmetric functions $Q$ were classified in \cite{ABS} under
the following additional assumption.
\begin{itemize}
\item \textit{Tetrahedron property}: The value $x_{123}$, which is
well defined due to 3D consistency, depends on $x_1$, $x_2$, and
$x_3$, but not on $x$.
\end{itemize}
The classification of equations (\ref{system}) up to M\"{o}bius
transformation results in a list (the so-called ABS list) of 9
canonical equations, named Q1--Q4, H1--H3, and A1-A2. The a priori
assumption of the tetrahedron property was replaced with certain
non-degeneracy conditions in \cite{ABS2}. This leads to the list
Q1--Q4.

An important device used for the classification are the
biquadratic polynomials $h$ and $g$ associated with the edges and
diagonals of the elementary quadrilateral, respectively. They are
obtained from $Q$ by discriminant-like operations eliminating two
of the four variables. For instance,
\[
QQ_{yv}-Q_yQ_v=k(\a,\b)h(x,u;\a), \quad
QQ_{yu}-Q_yQ_u=k(\a,\b)h(x,v;\b),
\]
\[
QQ_{uv}-Q_uQ_v=k(\a,\b)g(x,y;\a-\b).
\]
Here, the subscripts denote partial derivatives, and
$k(\a,\b)=-k(\b,\a)$ is a normalizing factor that makes each edge
polynomial $h$ depend only on the parameter assigned to the
corresponding edge. The polynomials $g$ associated with the
diagonals depend only on the difference $\a-\b$ for a suitable
choice of parameters, which are naturally defined up to
simultaneous re-parametrization $\alpha\mapsto \rho(\alpha)$,
$\beta\mapsto\rho(\beta)$.

The following lemma will be instrumental in the proof of our main
result.
\begin{lemma}\label{Lemma biquadr}
For any quad-equation from the ABS list, the following identity is
satisfied for solutions of $Q=0$:
\begin{equation}\label{biquad id}
h(x,u;\a)h(y,v;\a)=h(x,v;\b)h(y,u;\b)=g(x,y;\a-\b)g(u,v;\a-\b).
\end{equation}
\end{lemma}
 We refer the reader to \cite{ABS} for further details and
a proof of Lemma \ref{Lemma biquadr}.

\section{Three-leg forms}
\label{sect: 3leg}

Equation (\ref{system}) is said to possess a {\em three-leg form}
centered at $x$ if it is equivalent to the equation
\begin{equation}\label{eq:3leg}
    \psi(x,u;\a)-\psi(x,v;\b)=\phi(x,y;\a-\b),
\end{equation}
for some functions $\psi$ and $\phi$, see Fig.~\ref{Fig: 3leg}. It
follows that the function $\phi$ must be odd with respect to the
parameter: $\phi(x,y;-\gamma)=-\phi(x,y;\gamma)$. It turns out
\cite{ABS} that all equations from the ABS list possess three-leg
forms. Moreover, an examination of the list of three-leg forms
leads to the following
\begin{itemize}
\item \textit{Observation}: For equations Q1--Q4, the functions
corresponding to the ``short'' and to the ``long'' legs coincide:
$\psi(x,u;\a)=\phi(x,u;\a)$. Each equation H1--H3 and A1--A2
shares the ``long'' leg function $\phi(x,y;\a-\b)$ with some of
the equations Q1--Q3, but has a different ``short'' legs function
$\psi(x,u;\a)$.
\end{itemize}
\noindent There are many applications of the three-leg form.

First, let $\mathcal B$ be the ``black'' subgraph of the bipartite
quad-graph $\mathcal D$ on which the system of integrable
quad-equations is considered. The edges of $\mathcal B$ are the
diagonals of the quadrilateral faces of $\mathcal D$ connecting
the ``black'' pairs of vertices. Let the pairs of labels be
assigned to the edges of $\mathcal B$ according to Fig.\,\ref{Fig:
3leg}, so that $(\a,\b)$ is assigned to the edge $(x,y)$. Then the
restriction of any solution of the system of quad-equations to the
set of ``black'' vertices satisfies the so called Laplace type
equations. For $x\in V(\mathcal B)$ such an equation reads:
\begin{equation}\label{Toda}
  \sum_{(x,y_k)\in E(\mathcal B)}\phi(x,y_{k};\a_k-\a_{k+1})=0.
\end{equation}
Here, the sum is taken over all edges $(x,y_k)$ of $\mathcal B$
incident with $x$ in counterclockwise order, and $(\a_k,\a_{k+1})$
are the corresponding pairs of parameters. Equation (\ref{Toda})
is derived by adding the three-leg forms of the quad-equations for
all quadrilaterals of $\mathcal D$ adjacent to $x$, where the
contributions from the ``short'' legs cancel out. Of course,
similar Laplace type equations hold also for the ``white''
subgraph of $\mathcal D$.

Another application of the three-leg form is the derivation of the
tetrahedron property. Adding the tree-leg equations centered at
$x_{123}$ on the three faces of the 3D cube adjacent to $x_{123}$
leads to the equation
\begin{equation}\label{eq:tetr}
\phi(x_{123},x_1;\a_2-\a_3)+\phi(x_{123},x_2;\a_3-\a_1)+
\phi(x_{123},x_3;\a_1-\a_2)=0,
\end{equation}
which relates the fields at the vertices of the ``white''
tetrahedron in Fig.~\ref{3D}. According to the above observation,
this equation is actually equivalent to
\begin{equation}\label{eq:tetr Q}
\widehat{Q}(x_{123},x_1,x_2,x_3;\a_2-\a_3,\a_2-\a_1)=0,
\end{equation}
where the function $\widehat{Q}(x,u,y,v;\a,\b)$ is multi-affine,
and, moreover, always belongs to the list Q1--Q4 (it plainly
coincides with $Q$ for any of the equations Q1--Q4).

Not only does the existence of the three-leg form yield the
tetrahedron property of the 3D consistent equations. The converse
is also true: it has been proved in \cite{AS} that $D_4$ symmetry
and the existence of a three-leg form imply 3D consistency.

\section{Lagrangian structures}\label{sect:L}

We use the following technical statement to establish the
Lagrangian structure of 3D consistent equations \cite{ABS}.
\begin{lemma}\label{Lemma L}
For any quad-equation from the ABS list, there exists a change of
variables, $x=f(X)$, $u=f(U)$, etc., such that in the new
variables  the leg functions $\psi$ and $\phi$ possess
antiderivatives with respect to the first argument $X$ that are
symmetric with respect to the permutation $X\leftrightarrow U$ and
$X\leftrightarrow Y$, respectively. In other words, there exist
functions $L(X,U;\a)=L(U,X;\a)$ and
$\Lambda(X,Y;\a-\b)=\Lambda(Y,X;\a-\b)$ such that
\begin{align}
\psi(x,u;\a)=\psi(f(X),f(U);\a) & =  \frac{\partial}{\partial
X}L(X,U;\a),
\label{Psi}\\
\phi(x,y;\a-\b)=\phi(f(X),f(Y);\a-\b) & =
\frac{\partial}{\partial X} \Lambda(X,Y;\a-\b).\label{Phi}
\end{align}
\end{lemma}
This follows from the easily verified fact that the derivatives of
the leg functions with respect to their second argument,
$\partial\psi/\partial U$ and $\partial\phi/\partial Y$, are
symmetric with respect to $X\leftrightarrow U$ and
$X\leftrightarrow Y$, respectively. Lemma \ref{Lemma L} has the
following corollaries \cite{ABS}.
\begin{proposition}\label{prop:Toda Lagr}
For any quad-equation from the ABS list on a bipartite quad-graph
$\mathcal D$, the corresponding Laplace type equations
(\ref{Toda}) on the ``black'' subgraph $\mathcal B$ are the
Euler-Lagrange equations for the action functional
\begin{equation}\label{Toda action}
S_{\mathcal B}=\sum_{(x,y)\in E(\mathcal B)} \Lambda(X,Y;\a-\b),
\end{equation}
where the pairs of parameters $(\a,\b)$ are assigned to the
``black'' edges $(x,y)$ as in Fig.~\ref{Fig: 3leg}.
\end{proposition}
\begin{proposition}\label{prop: quad Lagr}
For any quad-equation from the ABS list on the regular square
lattice $\mathbb Z^2$, the solutions are critical points of the
functional
\begin{equation}\label{action ABS}
    S=\sum_{(x,x_1)\in E_1}L(X,X_1;\alpha_1) - \sum_{(x,x_2)\in
    E_2}L(X,X_2;\alpha_2) - \sum_{(x_1,x_2)\in
    E_3}\Lambda(X_1,X_2;\alpha_1-\alpha_2),
\end{equation}
where $E_1$ and $E_2$ denote the set of horizontal and vertical
edges of the square lattice $\mathbb{Z}^2$, and $E_3$ denotes the
set of diagonals of all elementary quadrilaterals from north-west
to south-east.
\end{proposition}
The proof of Proposition \ref{prop:Toda Lagr} is obvious, the
proof of Proposition \ref{prop: quad Lagr} is based on the fact
that $\partial S/\partial X$ is the sum of the three-leg equations
on two squares adjacent to $x$ (to the north-west and to the
south-east of $x$).

\section{Fundamental property of Lagrangians on a \\ single quad}
\label{sect:one quad}

\begin{theorem}\label{th:one quad}
For any equation from the ABS list, considered on a single
quadrilateral, the Lagrangians $L,\Lambda$ can be chosen so that
the following relation holds if equation (\ref{system}) is
satisfied:
\begin{align}\label{for one quad}
& L(X,U;\a)+L(Y,V;\a)-L(X,V;\b)-L(Y,U;\b)\nonumber\\
& -\Lambda(X,Y;\a-\b)-\Lambda(U,V;\a-\b)=0.
\end{align}
\end{theorem}
\noindent {\bf Proof.} Since the symmetric antiderivatives $L$ and
$\Lambda$ are determined only up to constant terms (depending on
the corresponding parameters), the theorem is actually equivalent
of  to the statement that for any choice of $L$, $\Lambda$ there
holds (for solutions of $Q=0$):
\begin{equation}\label{aux}
\Theta=\rho(\a)-\rho(\b)-\sigma(\a-\b),
\end{equation}
where $\Theta$ stands for the left-hand side of (\ref{for one
quad}), and $\rho$, $\sigma$ are some functions depending  only on
the parameters, as indicated by the notation.

To show that the function $\Theta=\Theta(X,U,Y,V)$ is constant on
the three-dimensional manifold in $(\mathbb C\mathbb P^1)^4$
consisting of solutions of $Q(x,u,y,v;\a,\b)=0$, it is enough to
prove that the directional derivatives of $\Theta$ along all
tangent vectors of this manifold vanish. We prove a stronger
claim, namely that the gradient of $\Theta$ vanishes on this
manifold. This claim is an immediate consequence of the existence
of the three-leg equations centered at each vertex of the
elementary quad. Indeed, by virtue of (\ref{Psi}), (\ref{Phi}),
and (\ref{eq:3leg}), one has:
\begin{equation}
\frac{\partial\Theta}{\partial
X}=\psi(x,u;\a)-\psi(x,v;\b)-\phi(x,y;\a-\b)=0.
\end{equation}
Similarly, one shows that $\partial\Theta/\partial
Y=\partial\Theta/\partial U=\partial\Theta/\partial V=0$ for
solutions. It remains to show that the constant value of $\Theta$
is of the form (\ref{aux}). The proof of this fact is based on
identity (\ref{biquad id}) and the following lemma.
%This is trivially verified for equation (Q1)$_{\d=0}$, for which
%one can take
%\[
%L(x,u;\a)=\a\log|x-u|,\quad \Lambda(x,y;\a-\b)=(\a-\b)\log|x-y|,
%\] and then
%\begin{eqnarray*}
%\Theta & =& \a\log\left|\frac{(x-u)(y-v)}{(x-y)(u-v)}\right|-
%\b\log\left|\frac{(x-v)(y-u)}{(x-y)(u-v)}\right|\\
% & = &
%%2\a\log\left|\frac{\a}{\a-\b}\right|-2\b\log\left|\frac{\b}{\a-\b}\right|\\
% & = & 2\a\log|\a|-2\b\log|\b|-2(\a-\b)\log|\a-\b|.
%\end{eqnarray*}
%Analogously, the claim is directly checked for equation (H1), for
%which one can take
%\[
%L(x,u;\a)=\frac{1}{2}(x+u)^2,\quad
%\Lambda(x,y;\a-\b)=(\a-\b)\log|x-y|,
%\]
%and then
%\begin{eqnarray*}
%\Theta & =& (x-y)(u-v)-(\a-\b)\log|(x-y)(u-v)|\\
% & = & \a-\b-(\a-\b)\log|\a-\b|.
%\end{eqnarray*}

\begin{lemma}\label{Lemma dL/da}
For any equation from the ABS list, we have:
\begin{eqnarray}
\frac{\partial L(X,U;\a)}{\partial\a} & = & \log
h(x,u;\a)+\kappa(X)+\kappa(U)+c(\a),\label{eq: dL/da}\\
\frac{\partial \Lambda(X,Y;\a-\b)}{\partial\a} & = & \log
g(x,y;\a-\b)+\kappa(X)+\kappa(Y)+\g(\a-\b),\qquad \label{eq:
dLambda/da}
\end{eqnarray}
with certain functions $\kappa$, $c$, $\g$ depending only on the
indicated variables.
\end{lemma}
{\bf Proof.} Verify the relations obtained from (\ref{eq: dL/da}),
(\ref{eq: dLambda/da}) by differentiation with respect to $X$:
\[
\frac{\partial \psi(x,u;\a)}{\partial \a}=\frac{\partial}{\partial
X} \log h(x,u;\a)+\kappa'(X),
\]
\[
\frac{\partial \phi(x,y;\a-\b)}{\partial
\a}=\frac{\partial}{\partial X} \log g(x,y;\a-\b)+\kappa'(X).
\]
This can be done case by case, by a direct and simple check; the
leg functions $\psi$, $\phi$ and the polynomials $h$, $g$ are
given for all equations of the ABS list in the Appendix. Then
equations (\ref{eq: dL/da}), (\ref{eq: dLambda/da}) follow, since
both sides of each are symmetric with respect to $x\leftrightarrow
u$ and $x\leftrightarrow y$, respectively, and are defined up to
an additive function of $\a$, resp. of $\a-\b$. $\Box$

Lemma \ref{Lemma dL/da} and identity (\ref{biquad id}) imply
\[
\frac{\partial \Theta}{\partial\a}= 2c(\a)-2\gamma(\a-\b), \quad
\frac{\partial \Theta}{\partial\b}= -2c(\b)+2\gamma(\a-\b),
\]
which yields (\ref{aux}). This completes the proof of Theorem
\ref{th:one quad}. $\Box$
%Analogously, for equation (H2) we find:
%\[
%\frac{\partial L(X,U;\a)}{\partial\a}=\log h(x,u;\a),\quad
%\frac{\partial
%\Lambda(X,Y;\a)}{\partial\a}=\log\Big(2(\a-\b)^2g(x,y;\a-\b)\Big),
%\]
%while for equation (H3):
%\begin{eqnarray*}
%\frac{\partial L(X,U;\a)}{\partial\a} & = & X+U-\log h(x,u;\a),\\
%\frac{\partial \Lambda(X,Y;\a)}{\partial\a} & = & X+Y
%-\log\Big(\frac{\sinh (\a-\b)}{2}g(x,y;\a-\b)\Big).
%\end{eqnarray*}
%In the same way as above, there follows that $\partial
%\Theta/\partial\a=-\partial\Theta/\partial\b=2\log \rho(\a-\b)$.

\section{Flip invariance of the action functionals}\label{sect: flip}

\begin{figure}[htbp]
\setlength{\unitlength}{0.08em}
\begin{center}
\begin{picture}(350,130)(-30,0)
  \dottedline{3}(0,25)(40,0)(80,25)(80,75)(40,100)(0,75)(0,25)
  \dottedline{3}(40,0)(40,50)
  \dottedline{3}(40,50)(0,75)
  \dottedline{3}(40,50)(80,75)
  \put(50,50){$x$}
  \put(40,-10){$x_1$}
  \put(87,27){$x_{12}$}
  \put(85,80){$x_2$}
  \put(30,110){$x_{23}$}
  \put(-15,80){$x_3$}
  \put(-22,30){$x_{13}$}
  \put(40,50){$\circle*{6}$}
  \put(40,100){$\circle*{6}$}
  \put(80,25){$\circle*{6}$}
  \put(0,25){$\circle*{6}$}
  {\thicklines\path(40,50)(40,100)\path(40,50)(80,25)\path(40,50)(0,25)}
  \put(130,50){\vector(1,0){20}}
  \dottedline{3}(200,25)(240,0)(280,25)(280,75)(240,100)(200,75)(200,25)
  \dottedline{3}(240,50)(280,25)
  \dottedline{3}(240,50)(200,25)
  \dottedline{3}(240,50)(240,100)
  \put(230,40){$x_{123}$}
  \put(240,-10){$x_1$}
  \put(287,27){$x_{12}$}
  \put(285,80){$x_2$}
  \put(230,110){$x_{23}$}
  \put(185,80){$x_3$}
  \put(178,30){$x_{13}$}
  \put(240,100){$\circle*{6}$}
  \put(280,25){$\circle*{6}$}
  \put(200,25){$\circle*{6}$}
  {\thicklines\path(240,100)(280,25)(200,25)(240,100)}
\end{picture}
\end{center}
\caption{Star-triangle flip.} \label{Fig:flip}
\end{figure}
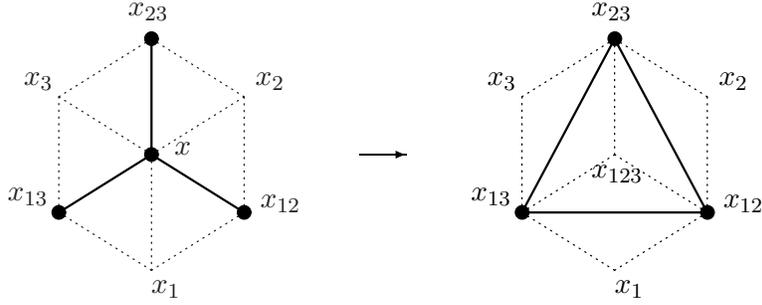

The following theorem establishes the flip invariance for the
discrete Laplace type systems (with the Lagrangian structure
described in Proposition \ref{prop:Toda Lagr}).
\begin{theorem}\label{th:star-triang}
The Lagrangian $\Lambda$ for a discrete Laplace type system that
comes from an equation of the ABS list can be chosen so that the
following star-triangle relation is satisfied for solutions:
\begin{align}\label{Lagr flip star-triang}
&
\Lambda(X,X_{12};\alpha_1-\alpha_2)+\Lambda(X,X_{23};\alpha_2-\alpha_3)+
\Lambda(X,X_{13};\alpha_3-\alpha_1)\nonumber\\
& +\Lambda(X_{23},X_{13};\alpha_1-\alpha_2)+
\Lambda(X_{13},X_{12};\alpha_2-\alpha_3)+
\Lambda(X_{12},X_{23};\alpha_3-\alpha_1)=0,\quad
\end{align}
see Fig.~\ref{Fig:flip}.
\end{theorem}
%-----------------------------------------------------------------
%-----------------------------------------------------------------

\begin{figure}[htbp]
\setlength{\unitlength}{0.08em}
\begin{center}
\begin{picture}(350,100)(-30,30)
  \put(50,50){$x$}
  \put(87,27){$x_{12}$}
  \put(30,110){$x_{23}$}
  \put(-22,30){$x_{13}$}
  \put(40,50){$\circle*{6}$}
  \put(40,100){$\circle*{6}$}
  \put(80,25){$\circle*{6}$}
  \put(0,25){$\circle*{6}$}
  {\thicklines\path(40,50)(40,100)\path(40,50)(80,25)\path(40,50)(0,25)
  \path(40,100)(80,25)(0,25)(40,100)}
  \put(185,20){$x$}
  \put(290,20){$u$}
  \put(290,110){$y$}
  \put(185,110){$v$}
  \put(280,105){$\circle*{6}$}
  \put(200,105){$\circle*{6}$}
  \put(280,25){$\circle*{6}$}
  \put(200,25){$\circle*{6}$}
  {\thicklines\path(200,25)(280,25)(280,105)(200,105)(200,25)(280,105)
              \path(280,25)(200,105)}
\end{picture}
\end{center}
\caption{A tetrahedron vs. a quadrilateral with diagonals}
\label{Fig:tetr-quad}
\end{figure}
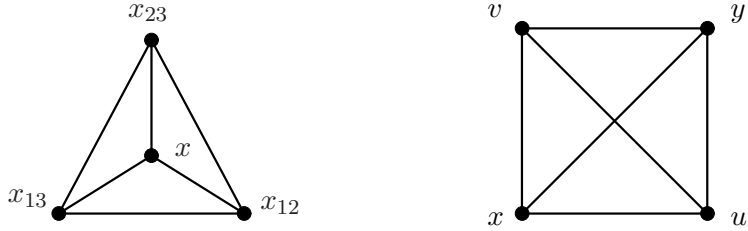

{\bf Proof.} Formula (\ref{Lagr flip star-triang}) involves the
four black points $x$, $x_{12}$, $x_{23}$, $x_{13}$, which are
related by a multi-affine equation
\[
\widehat{Q}(x,x_{12},x_{23},x_{13};\a_1-\a_2,\a_1-\a_3)=0,
\]
which belongs to the list Q1--Q4, compare with (\ref{eq:tetr Q}).
Therefore, the claim is a particular case of Theorem \ref{th:one
quad}. Indeed, combinatorially a tetrahedron is not different from
a quadrilateral with diagonals, see Fig.~\ref{Fig:tetr-quad}.
$\Box$
%-----------------------------------------------------------------
%-----------------------------------------------------------------

Such a statement was previously established in \cite{BMS} for the
discrete Laplace type system which describes the radii of circle
patterns with prescribed intersection angles and which comes from
the so called Hirota system, a version of (H3)$_{\delta=0}$. In
that paper, the action functional is derived as a classical limit
of the partition function of the so called quantum Faddeeev-Volkov
model. The corresponding property of the quantum model is the
famous Yang-Baxter relation, the invariance of the partition
function under a star-triangle transformation of the Boltzmann
weights. The corresponding classical result is also established in
\cite{BMS}, by direct computations involving the dilogarithm
function.

The flips described by Theorem \ref{th:star-triang} can be
considered as elementary transformations either of a planar
quad-graph, or, alternatively, of its realization as a
quad-surface in a multidimensional square lattice $\mathbb Z^m$.
The Lagrangian formulation of quad-equations on $\mathbb Z^m$ is
the main subject of \cite{Lobb}.

The Lagrangian formulation of systems on $\mathbb Z^2$ used in
\cite{Lobb} is
\begin{equation}\label{action LN}
    S=\sum_{\mathbb{Z}^2}\mathcal{L}(X,X_1,X_2;\alpha_1,\alpha_2),
\end{equation}
where the 3-point Lagrangian $\mathcal{L}$ should be interpreted
as a discrete 2-form, i.e., a real-valued function defined on
oriented elementary squares and changing sign upon changing the
orientation of the square. It is easily seen that the sum
(\ref{action LN}) is nothing but a re-arrangement of the sum
(\ref{action ABS}), with
\begin{equation}\label{lag LN}
\mathcal{L}(X,X_1,X_2;\alpha_1,\alpha_2)=
L(X,X_1;\alpha_1)-L(X,X_2;\alpha_2)-\Lambda(X_1,X_2;\alpha_1-\alpha_2).
\end{equation}

The main idea of the paper \cite{Lobb} is to extend the functional
(\ref{action LN}) to quad-surfaces $\Sigma$ in the
multidimensional square lattice according to the formula
\begin{equation}
S=\sum_{\sigma_{ij}\in\Sigma}\mathcal L(\sigma_{ij})
\end{equation}
where for each elementary square
$\sigma_{ij}=(n,n+e_i,n+e_i+e_j,n+e_j)$ there holds
\begin{eqnarray}\label{lag LN ij}
\lefteqn{\mathcal
L(\sigma_{ij})=\mathcal{L}(X,X_i,X_j;\alpha_i,\alpha_j)}\nonumber\\
&&=L(X,X_i;\alpha_i)-
L(X,X_j;\alpha_j)-\Lambda(X_i,X_j;\alpha_i-\alpha_j).
\end{eqnarray}
 Let $\Delta_i$ denote the difference operator that acts on vertex
functions, $\Delta_if(x)=f(x_i)-f(x)$, so that, e.g.,
$\Delta_if(x,x_j,x_k)=f(x_i,x_{ij},x_{ik})-f(x,x_j,x_k)$.

\begin{theorem}\label{th: LN}
For any system of quad-equations from the ABS list on $\mathbb
Z^m$, the Lagrangian $\mathcal L$ given by (\ref{lag LN ij})
satisfies the following relation for solutions:
\begin{equation}\label{closure relation}
\Delta_1\mathcal{L}(X,X_2,X_3;\alpha_2,\alpha_3)+
\Delta_2\mathcal{L}(X,X_3,X_1;\alpha_3,\alpha_1)+
\Delta_3\mathcal{L}(X,X_1,X_2;\alpha_1,\alpha_2)=0.
\end{equation}
\end{theorem}
\noindent This means that the value of the action functional {\em
for a solution} remains invariant under flips of the quad-surface.
For some equations of the ABS list, namely for equations A1--A2,
H1--H3, Q1, (Q3)$_{\delta=0}$, Theorem \ref{th: LN} was proved in
\cite{Lobb} by long computations.

{\bf Proof of Theorem~\ref{th: LN}.} It is enough to combine the
statements of Theorem \ref{th:one quad} for the three
quadrilaterals adjacent to the vertex $x$ and the statement of
Theorem \ref{th:star-triang} for the black tetrahedron. $\Box$
\smallskip

The following alternative proof of Theorem \ref{th: LN}, not
relying on Theorem \ref{th:one quad}, is based on the same idea as
the proof of Theorem \ref{th:one quad} but is much easier. The
previous analysis of the constant value (\ref{aux}) is replaced by
a simple and case-independent argument.

{\bf Second proof of Theorem \ref{th: LN}.}  Let $\Delta$ denote
the expression on the left-hand side of (\ref{closure relation}),
considered as a function of 8 variables $x,x_i,x_{ij},x_{123}$. We
are going to show that $\Delta$ is constant on the manifold
$\mathcal{S}\subset (\mathbb C\mathbb P^1)^8$ of solutions of the
system of quad-equations on the 3D cube. This manifold is
four-dimensional and is parametrized, e.g., by $(x,x_1,x_2,x_3)$.
We want to show that the derivatives of $\Delta$ tangent to
$\mathcal S$ vanish. It turns out that a stronger property is
easier to show, namely, that ${\rm grad}\,\Delta=0$ on $\mathcal
S$.

\begin{figure}[htbp]
\begin{center}
\setlength{\unitlength}{0.05em}
\begin{picture}(200,250)(0,-10)
 \put(0,0){\circle*{10}}    \put(150,0){\circle{10}}
 \put(0,150){\circle{10}}   \put(150,150){\circle*{10}}
 \put(50,200){\circle*{10}} \put(200,200){\circle{10}}
 \put(50,50){\circle{10}}   \put(200,50){\circle*{10}}
 {\allinethickness{0.65mm}\path(150,5)(150,150)(5,150)
 \path(3.53,153.53)(50,200)(150,150)(200,50)(153.53,3.53)
 \path(3.53,146.47)(146.47,3.53)
 \path(1,145.5)(46.5,54)
 \path(145.5,1)(54,46.5)
 \dashline[100]{15}(50,200)(200,50)
 \dashline[80]{15}(50,55)(50,200)
 \dashline[80]{15}(55,50)(200,50)}
 {\thinlines\path(0,0)(46.47,46.47)
 \path(0,145)(0,0)(145,0)
 \path(150,150)(196.47,196.47)
 \path(50,200)(195,200)
 \path(200,195)(200,50)}
 %\dashline{3}(150,0)(0,150)
 %\dashline{3}(150,0)(70,50)
 %\dashline{3}(70,50)(0,150)
 %\dashline{3}(150,0)(220,200)
 %\dashline{3}(0,150)(220,200)
 %\dashline{3}(70,50)(220,200)
 \put(-20,-15){$x$} \put(-30,145){$x_3$}
 \put(160,-15){$x_1$} \put(162,140){$x_{13}$}
 \put(215,45){$x_{12}$} \put(210,210){$x_{123}$}
 \put(40,25){$x_2$}  \put(35,213){$x_{23}$}
\end{picture}
\caption{Octahedron}\label{octahedron}
\end{center}
\end{figure}
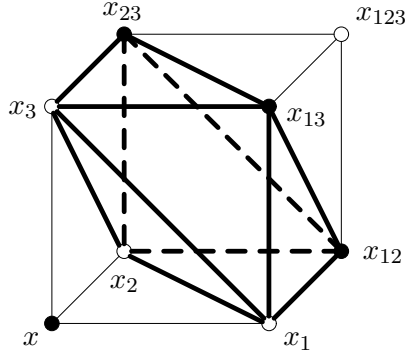

By the definition of the Lagrangian (\ref{lag LN ij}), we have:
\begin{eqnarray}\label{Theta}
\Delta & = & L(X_1,X_{12};\alpha_2)+L(X_2,X_{23};\alpha_3)
             +L(X_3,X_{13};\alpha_1)\nonumber\\
       &   & -L(X_1,X_{13};\alpha_3)-L(X_2,X_{12};\alpha_1)
             -L(X_3,X_{23};\alpha_2)\nonumber\\
       &   & -\Lambda(X_{12},X_{13};\alpha_2-\alpha_3)
             -\Lambda(X_{23},X_{12};\alpha_3-\alpha_1)
             -\Lambda(X_{13},X_{23};\alpha_1-\alpha_2)\nonumber\\
       &   & +\Lambda(X_{2},X_{3};\alpha_2-\alpha_3)+
             \Lambda(X_{3},X_{1};\alpha_3-\alpha_1)+
             \Lambda(X_{1},X_{2};\alpha_1-\alpha_2).\qquad
\end{eqnarray}
Thus, $\Delta$ does not depend on either $x$ or $x_{123}$, so that
its  domain of definition is better visualized as an octahedron as
shown in Fig.~\ref{octahedron} rather than an elementary cube, as
the original definition suggests.
%-----------------------------------------------------------------
%-----------------------------------------------------------------
It remains to show that $\Delta$ does not depend on $x_i$ and
$x_{ij}$ for solutions of the system of quad-equations. To show
that $\Delta$ does not depend on $x_1$, say, we compute, with the
help of (\ref{Psi}) and (\ref{Phi}):
\begin{equation*}
    \frac{\partial\Delta}{\partial X_1}=
    \psi(x_1,x_{12};\a_2)-\psi(x_1,x_{13};\a_3)+
    \phi(x_{1},x_{3};\a_3-\a_1)+\phi(x_{1},x_{2};\a_1-\a_2).
\end{equation*}
But the tree-leg forms of the quad-equations on the faces
$(x,x_1,x_{13},x_3)$ and $(x,x_1,x_{12},x_2)$, centered at $x_1$
are
\begin{eqnarray*}
    \psi(x_1,x_{13};\a_3)-\psi(x_1,x;\a_1)-\phi(x_1,x_3;\a_3-\a_1)=0,\\
    \psi(x_1,x_{12};\alpha_2)-
    \psi(x_1, x;\alpha_1)+\phi(x_{1},x_{2};\alpha_1-\alpha_2)=0.
\end{eqnarray*}
Therefore, for solutions we have $\partial\Delta/\partial X_1=0$.
That the partial derivatives of $\Delta$ with respect to all other
$x_i$ and $x_{ij}$ vanish is shown similarly, because all
variables enter symmetrically in $\Delta$. It is easy to
understand that the manifold of solutions $\mathcal S$ is a
connected algebraic manifold. Indeed, $\mathcal S =
(\mathbb{CP}^1)^4\setminus \tilde{\mathcal S}$, where $
\tilde{\mathcal S}$ consists of singular curves and therefore has
codimension two. Since ${\rm grad}\,\Delta=0$ on the connected
algebraic manifold $\mathcal S$, the function $\Delta$ is constant
on $\mathcal S$. It remains to show that the value of this
constant is $0$. We need only to compute $\Delta$ on a particular
solution. Consider a family of solutions defined by the following
conditions:
\begin{equation}\label{particular solution}
    x_{1}=x_{23}, \quad x_{2}=x_{13},\quad x_{3}=x_{12}.
\end{equation}
(We are grateful to K. Zuev for the suggestion to consider this
family.) Equations on the faces adjacent to the vertex $x$ give
three different expressions for $x$. Setting them equal means
imposing two (rational) conditions on the three initial values
$x_1,x_2,x_3$. Thus, there is a one-parameter family of solutions
satisfying (\ref{particular solution}). Thanks to the symmetry of
$L$ and $\Lambda$ one sees immediately from (\ref{Theta}) that
$\Delta=0$ on any solution from the family (\ref{particular
solution}). This finishes the proof of Theorem \ref{th: LN}.
$\Box$

\section{Appendix: ABS list}
{\bf List Q:}

\begin{itemize}
\item[{\rm(Q1)}$_{\delta=0}$:] $\quad
Q=\alpha(xu+yv)-\beta(xv+yu)-
  (\alpha-\beta)(xy+uv),$
  \item[] $\quad\psi(x,u;\alpha)=\dfrac{\alpha}{x-u}$,
  \item[] $\quad h(x,u;\alpha)=\dfrac{1}{2\alpha}(x-u)^2$;

\item[{\rm(Q1)}$_{\delta=1}$:] $\quad
Q=\alpha(xu+yv)-\beta(xv+yu)-
  (\alpha-\beta)(xy+uv)+\alpha\beta(\alpha-\beta),$

\item[]
$\quad\psi(x,u;\alpha)=\log\dfrac{x-u+\alpha}{x-u-\alpha},$

\item[]  $\quad
h(x,u;\alpha)=\dfrac{1}{2\alpha}\Big((x-u)^2-\a^2\Big)=
\dfrac{1}{2\alpha}(x-u+\a)(x-u-\a);$

\item[{\rm(Q2)}:]
 $\quad Q=\alpha(xu+yv)-\beta(xv+yu)-(\alpha-\beta)(xy+uv)$

 $\qquad\qquad +\alpha\beta(\alpha-\beta)(x+u+y+v)
           -\alpha\beta(\alpha-\beta)(\alpha^2-\alpha\beta+\beta^2),$

\item[] $\quad x=X^2$,

\item[]
$\quad\psi(x,u;\alpha)=\log\dfrac{(X+U+\a)(X-U+\a)}{(X+U-\a)(X-U-\a)}$,

\item[] $\quad
h(x,u;\alpha)=\dfrac{1}{4\a}\Big((x-u)^2-2\alpha^2(x+u)+\a^4\Big)$

\item[] $\qquad=\dfrac{1}{4\a}(X+U+\a)(X-U+\a)(X+U-\a)(X-U-\a);$
\end{itemize}

\begin{itemize}
\item[{\rm(Q3)}$_{\d=0}$:]
 $\quad Q=\sin(\alpha)(xu+yv)-\sin(\beta)(xv+yu)-\sin(\alpha-\beta)(xy+uv),$

\item[] $\quad x=\exp(iX)$,

\item[]
$\quad\psi(x,u;\alpha)=\log\dfrac{\sin\!\left(\dfrac{X-U+\a}{2}\right)}
{\sin\!\left(\dfrac{X-U-\a}{2}\right)}$,

\item[] $\quad
h(x,u;\alpha)=\dfrac{1}{\sin(\a)}\Big(x^2+u^2-2\cos(\a)xu\Big)$

\item[]
$\qquad=\dfrac{\exp(iX)\exp(iU)}{\sin(\a)}\sin\!\left(\dfrac{X-U+\a}{2}\right)
\sin\!\left(\dfrac{X-U-\a}{2}\right)$;

\item[{\rm(Q3)}$_{\d=1}$:]
 $\quad Q=\sin(\alpha)(xu+yv)-\sin(\beta)(xv+yu)
      -\sin(\alpha-\beta)(xy+uv)$

 $\qquad\qquad +\sin(\alpha-\beta)\sin(\alpha)\sin(\beta),$

\item[] $\quad x=\sin(X)$,

\item[]
$\quad\psi(x,u;\alpha)=\log\dfrac{\cos\!\left(\dfrac{X+U+\a}{2}\right)
\sin\!\left(\dfrac{X-U+\a}{2}\right)}{\cos\!\left(\dfrac{X+U-\a}{2}\right)
\sin\!\left(\dfrac{X-U-\a}{2}\right)}$,

\item[] $\quad
h(x,u;\alpha)=\dfrac{1}{2\sin(\a)}\Big(x^2+u^2-2\cos(\a)xu-\sin^2(\a)\Big)$

\item[] $\qquad
=\dfrac{2}{\sin(\a)}\cos\!\left(\dfrac{X+U+\a}{2}\right)
\cos\!\left(\dfrac{X+U-\a}{2}\right)
\sin\!\left(\dfrac{X-U+\a}{2}\right)
\sin\!\left(\dfrac{X-U-\a}{2}\right);$

\item[{\rm(Q4)}:]
        $\quad Q=\sn(\alpha)(xu+yv)-\sn(\beta)(xv+yu)
        -\sn(\alpha-\beta)(xy+uv)$

        $\qquad\qquad  +\sn(\alpha-\beta)\sn(\alpha)\sn(\beta)
        (1+k^2xuyv),$

\item[] $\quad x=\sn(X)$,

\item[] $\quad\psi(x,u;\alpha)=\log\dfrac
{\Theta_2\!\left(\dfrac{X+U+\a}{2}\right)
\Theta_3\!\left(\dfrac{X+U+\a}{2}\right)
\Theta_1\!\left(\dfrac{X-U+\a}{2}\right)
\Theta_4\!\left(\dfrac{X-U+\a}{2}\right)}
{\Theta_2\!\left(\dfrac{X+U-\a}{2}\right)
\Theta_3\!\left(\dfrac{X+U-\a}{2}\right)
\Theta_1\!\left(\dfrac{X-U-\a}{2}\right)
\Theta_4\!\left(\dfrac{X-U-\a}{2}\right)},$

\item[] $\quad
h(x,u;\alpha)=\dfrac{1}{2\,\sn(\a)}\Big(x^2+u^2-2\cn(\a)\dn(\a)xu-\sn^2(\a)-
k^2\sn^2(\a)x^2u^2\Big)$

\item[]
$\qquad=\dfrac{2\vartheta_4^2/\vartheta_2^4}{\sn(\a)}\cdot\dfrac{1}{\Theta_4^2(\a)
\Theta_4^2(X)\Theta_4^2(U)}$

\item[] $\qquad\times\Theta_2\!\left(\dfrac{X+U+\a}{2}\right)
\Theta_3\!\left(\dfrac{X+U+\a}{2}\right)
\Theta_1\!\left(\dfrac{X-U+\a}{2}\right)
\Theta_4\!\left(\dfrac{X-U+\a}{2}\right)$

\item[] $\qquad\times\Theta_2\!\left(\dfrac{X+U-\a}{2}\right)
\Theta_3\!\left(\dfrac{X+U-\a}{2}\right)
\Theta_1\!\left(\dfrac{X-U-\a}{2}\right)
\Theta_4\!\left(\dfrac{X-U-\a}{2}\right).$

  \end{itemize}
{\bf List H:}
\begin{itemize}
  \item[{\rm(H1)}] \quad $Q=(x-y)(u-v)+\b-\a$,

  \item[] $\quad \psi(x,u;\a)=x+u,\qquad
           \phi(x,y;\a-\b)=\dfrac{\a-\b}{x-y}$,

  \item[] $\quad h(x,u;\a)=1,\qquad
           g(x,y;\a-\b)=\dfrac{(x-y)^2}{\a-\b}$;

  \item[{\rm(H2)}] \quad $Q=(x-y)(u-v)+(\beta-\alpha)(x+u+y+v)+
                   \beta^2-\alpha^2$,
  \item[] $\quad \psi(x,u;\a)=\log(x+u+\a),\qquad
                 \phi(x,y;\a-\b)=\log\dfrac{x-y+\a-\b}{x-y-\a+\b}$,

  \item[] $\quad h(x,u;\a)=x+u+\a,\qquad
           g(x,y;\a-\b)=\dfrac{1}{2(\a-\b)}\Big((x-y)^2-(\a-\b)^2\Big)$;

  \item[{\rm(H3)}] \quad $Q=e^\a(xu+yv)-e^\b(xv+yu)+
                   \delta\left(e^{2\a}-e^{2\b}\right)$,

  \item[] $\quad x=e^X$,

  \item[] $\quad \psi(x,u;\a)=-\log(xu+\d e^\a)=
                 -\log\left(e^{X+U}+\d e^{\a}\right)$,

  \item[] $\quad \phi(x,y;\a-\b)=\log\dfrac{e^\a x-e^\b y}{e^\b x-e^\a y}=
             \log\dfrac{\sinh\!\left(\dfrac{X-Y+\a-\b}{2}\right)}
                       {\sinh\!\left(\dfrac{X-Y+\b-\a}{2}\right)}$,

  \item[] $\quad h(x,u;\a)=xu+\d e^\a=e^{X+U}+\d e^{\a}$,

  \item[] $\quad
       g(x,y;\a-\b)=\dfrac{1}{e^{2\a}-e^{2\b}}(e^\a x-e^\b y)(e^\b x-e^\a y)$

  \item[] $\qquad =\dfrac{2e^{X+Y}}{\sinh(\a-\b)}
               \sinh\!\left(\dfrac{X-Y+\a-\b}{2}\right)
               \sinh\!\left(\dfrac{X-Y+\b-\a}{2}\right)$;

\end{itemize}
{\bf List A:}
\begin{itemize}
\item[{\rm(A1)}$_{\d=0}$] \quad
  $Q=\a(xu+yv)-\b(xv+yu)+(\a-\b)(xy+uv)$,

\item[] $\quad\psi(x,u;\a)=\dfrac{\a}{x+u}$,\qquad
$\phi(x,y;\a-\b)=\dfrac{\a-\b}{x-y}$,

  \item[] $\quad h(x,u;\alpha)=\dfrac{1}{2\alpha}(x+u)^2$,\qquad
$g(x,y;\a-\b)=\dfrac{1}{2(\a-\b)}(x-y)^2$;

\item[{\rm(A1)}$_{\d=1}$] \quad
  $Q=\alpha(xu+yv)-\beta(xv+yu)+(\alpha-\beta)(xy+uv)
     -\alpha\beta(\alpha-\beta)$,

\item[] $\quad\psi(x,u;\a)=\log\dfrac{x+u+\a}{x+u-\a}$,\qquad
$\quad\phi(x,y;\a-\b)=\log\dfrac{x+y+\a-\b}{x+u-\a+\b}$,

\item[]  $\quad h(x,u;\a)=\dfrac{1}{2\a}\Big((x+u)^2-\a^2\Big)=
\dfrac{1}{2\alpha}(x+u+\a)(x+u-\a),$

\item[]  $\quad
g(x,y;\a-\b)=\dfrac{1}{2\a}\Big((x-y)^2-(\a-\b)^2\Big)=
\dfrac{1}{2\alpha}(x-y+\a-\b)(x-y-\a+\b);$

  \item[{\rm(A2)}]  \quad
  $Q=\sin(\alpha)(xv+yu)-\sin(\beta)(xu+yv)-\sin(\alpha-\beta)(1+xuyv)$,

\item[] $\quad x=\exp(iX)$,

\item[]
$\quad\psi(x,u;\alpha)=\log\dfrac{\sin\!\left(\dfrac{X+U+\a}{2}\right)}
{\sin\!\left(\dfrac{X+U-\a}{2}\right)}$,\quad
$\quad\phi(x,y;\a-\b)=\log\dfrac{\sin\!\left(\dfrac{X-Y+\a-\b}{2}\right)}
{\sin\!\left(\dfrac{X-Y-\a+\b}{2}\right)}$,

\item[] $\quad
h(x,u;\alpha)=-\dfrac{1}{\sin(\a)}\Big(x^2u^2+1-2\cos(\a)xu\Big)$

\item[]
$\qquad=\dfrac{\exp(iX)\exp(iU)}{\sin(\a)}\sin\!\left(\dfrac{X+U+\a}{2}\right)
\sin\!\left(\dfrac{X+U-\a}{2}\right)$,

\item[] $\quad
g(x,y;\a-\b)=\dfrac{1}{\sin(\a-\b)}\Big(x^2+y^2-2\cos(\a-\b)xy\Big)$

\item[]
$\qquad=\dfrac{\exp(iX)\exp(iY)}{\sin(\a-\b)}\sin\!\left(\dfrac{X-Y+\a-\b}{2}\right)
\sin\!\left(\dfrac{X-Y-\a+\b}{2}\right)$.

\end{itemize}

\end{document}